\documentclass[]{pasj02}
\usepackage[switch,mathlines]{lineno} 
\usepackage{ulem}
\usepackage{graphicx}

\jyear{2024}

\Received{}
\Accepted{}

\newcommand{\dm}{{\rm d}}
\newcommand{\msun}{\mathrm{M}_\odot}

\begin{document}

\title{The origin of the metallicity difference between star-forming and passive galaxies: Insights from $\nu^2$GC semi-analytic model}

\author{Qiansheng 
  \textsc{Liu},\altaffilmark{1}\altemailmark
  \email{liu@astro1.sci.hokudai.ac.jp}
 Takashi \textsc{Okamoto},\altaffilmark{2}\altemailmark\orcid{0000-0003-0137-2490} \email{takashi.okamoto@sci.hokudai.ac.jp}
 Taira \textsc{Oogi},\altaffilmark{3}
 and
 Masahiro \textsc{Nagashima}\altaffilmark{4}\orcid{0000-0003-2938-7096} 
}
\altaffiltext{1}{Department of Cosmosciences, Graduate School of Science, Hokkaido University, N10 W8 Kitaku, Sapporo 060-0810 Japan}
\altaffiltext{2}{Faculty of Science, Hokkaido University, N10 W8 Kitaku, Sapporo 060-0810 Japan}
\altaffiltext{3}{Department of Electrical and Computer Engineering, National Institute of Technology, Asahikawa College, Shunkodai 2-2-1-6, Asahikawa 071-8142 Japan}
\altaffiltext{4}{Faculty of Education, Bunkyo University, 3337, Minami-ogishima, Koshigaya, Saitama 343-8511 Japan}



\maketitle

\begin{abstract}
We investigate the origin of the observed metallicity difference between star-forming and passive galaxies using the semi-analytic galaxy formation model $\nu^2$GC. Our fiducial model successfully reproduces the observed metallicity differences in local galaxies while simultaneously matching the potential-metallicity relations of both star-forming and passive galaxies.
By varying the star formation efficiency, we identify strangulation as the primary driver of the metallicity difference. This finding highlights the critical role of star formation timescales in explaining the observed metallicity difference. 
Our results suggest that metallicity differences serve as a valuable diagnostic for evaluating star formation models in both semi-analytic models and cosmological simulations. Furthermore, galaxies quenched by processes resembling strangulation---where the supply of cold gas is halted in a slowly growing halo---exhibit higher metallicities than star-forming galaxies of the same stellar mass. In our model, this occurs in isolated, low-mass galaxies where rapid cooling leads to an effect resembling strangulation due to the discrete treatment of gas accretion onto dark matter halos. We propose that the metallicities of isolated, low-mass passive galaxies could provide key insights into refining models of hot gas halo growth.
\end{abstract}

\section{Introduction}

\noindent
Observationally, galaxies can be broadly classified into two categories {in several ways}: {star-forming and passive (or quiescent) galaxies based on their star formation rates \citep{noeske07, McGee2011, Wetzel2012}, blue and red based on colors \citep{Strateva2001, Blanton2003, Baldry2004, Wetzel2012, vanDerWel2014}, young and old based on ages \citep{Kauffmann2003, Gallazzi2008}, and late- and early-type based on morphologies \citep{Wuyts2011, vanDerWel2014}.
}
Understanding the mechanisms that shape this dichotomy is crucial for unraveling the complex processes governing galaxy evolution.

It is generally believed that a galaxy grows its stellar mass by forming stars at a rate which places it on the main-sequence of star-forming galaxies \citep{noeske07, speagle14}. When a galaxy is affected by some processes that `quench' the galaxy's star formation, it leaves the main-sequence and becomes a passive galaxy. Galaxy evolution predicted by cosmological simulations has supported this scenario (e.g.\cite{okamoto14, simba19, donnari19}).

Various mechanisms have been proposed to explain the quenching of star formation in galaxies, including ram-pressure stripping \citep{gunn72},  strangulation \citep{larson80}, feedback from active galactic nuclei (AGNs) in
massive halos \citep{croton06, bower06}, and halo quenching where gas from the intergalactic medium is shock-heated, becoming vulnerable to AGN feedback \citep{dekel06}. Mergers of galaxies can also be expected to play a role in quenching star formation by triggering starbursts and AGN activity \citep{cox08, simba19}.
Since different quenching processes operate on different timescales, observables that can discriminate the timescale can be used to identify the dominant quenching processes \citep{on04}.


\citet{peng15} found that, for local galaxies with stellar masses below $10^{11} \, \msun$, passive galaxies exhibit higher stellar metallicities than star-forming galaxies of the same stellar mass. They proposed that this metallicity difference could be explained by strangulation: a process where the supply of cold gas to a galaxy is halted, gradually ceasing star formation on a star formation timescale as the remaining cold gas is consumed by the star formation. Strangulation increases the stellar metallicity of quenched galaxies by preventing the dilution of the cold gas metallicity that would occur through the accretion of fresh, low-metallicity gas \citep{peng15}.
\citet{Trussler2020} further analyzed the stellar metallicities of star-forming, green valley, and passive galaxies, concluding that the combined effects of strangulation and outflows are primarily responsible for quenching most galaxies and explaining the metallicity differences observed between star-forming and passive galaxies.
On the other hand, \citet{vaughan22} argued that the relationship between stellar metallicity and the depth of the gravitational potential well, $\Phi$, is more fundamental than the mass--metallicity relation (MZR). 
They demonstrated that local galaxies follow a nearly universal relation between stellar metallicity and $\Phi$. 
Using a model where galaxy quenching probability depends on both mass and size, they successfully reproduced observed metallicity offsets under instantaneous quenching. 
They suggest that  the metallicity offset at a fixed stellar mass alone cannot serve as evidence for slow quenching processes.

Strangulation is expected to act on satellite galaxies. When a galaxy becomes a satellite, its hot halo gas is stripped off and becomes part of the hot atmosphere bound to the common halo, leaving denser cold gas {in the satellite}.
Since the hot gas in the common halo is too hot to be bound to the halo of the satellite, the satellite is no longer supplied with cold gas by gas cooling. The satellite becomes a passive galaxy as the cold gas is consumed by star formation. This process is explicitly modeled in semi-analytic (SA) galaxy formation models from the beginning of their development (e.g. \cite{kwg93, cole94}).
On the other hand, galaxy formation simulations should automatically include strangulation as a result of hydrodynamic interactions \citep{ofjt10}, although the timescale of hot gas stripping is not as short as usually assumed in SA models \citep{Kawata08, McCarthy2008}.
Thus, simulations and SA models should be able to reproduce the observed metallicity difference between star-forming and passive galaxies if its origin is strangulation.

However, \citet{okamoto17} found that the three independent cosmological simulations, the simulations by \citet{okamoto14}, the Illustris simulation (\cite{illustris}), and the EAGLE simulation (\cite{eagle}), all failed to reproduce the observed metallicity difference, predicting that both populations have almost the same metallicity for a given galaxy stellar mass. Similarly, \citet{fontanot21} showed that the stellar metallicities of local star-forming and passive galaxies are almost the same for a given galaxy stellar mass using their SA model. Therefore, the metallicity difference between star-forming and passive galaxies is still not understood in the context of galaxy formation.

In this paper, we test our SA model, $\nu^2$GC \citep{nu2GC, nu2GCagn},
which incorporates results from $N$-body simulations to follow the formation and evolution of {dark matter (DM)} halos and uses phenomenological equations to model the complicated baryon physics inside each {halo},
against the observed metallicity difference. We then investigate the origin of this metallicity difference to understand why the previous galaxy formation simulations and models failed to reproduce it.

This paper is organized as follows. In Section 2, we describe our SA galaxy formation model. In Section 3, we present our main results, including the MZRs for star-forming and passive galaxies, and comparisons between central and satellite galaxies. In Section 4, we discuss the implications of our results in the context of the strangulation hypothesis and other quenching mechanisms. Finally, in Section 5, we summarize our conclusions and suggest possible directions for future research.

\section{Model descriptions}\label{Model descriptions}
 \noindent
 In this study, we employ an SA model named $\nu^2$GC \citep{nu2GC, nu2GCagn}. The formation histories of DM halos (merger trees) are constructed by using cosmological $N$-body simulations, while galaxy formation processes in these halos, such as gas cooling, star formation, supernova feedback, growth of supermassive black holes (SMBHs), and feedback from active galactic nuclei (AGNs), are modeled by using phenomenological equations. This model is built upon large cosmological {$N$}-body simulations, designed to investigate the formation and evolution of galaxies in a cosmological context \citep{nu2GC}. 

One of the strengths of $\nu^2$GC is its ability to combine the large-scale dynamics of DM, tracked via $N$-body simulations, with semi-analytical prescriptions that model key baryonic processes. 
This model has been validated against diverse observational datasets. \citet{nu2GCagn} demonstrated its capability to accurately reproduce key galactic and SMBH scaling relations, including stellar mass functions for galaxies at $z < 4$, AGN luminosity functions at $z < 4$, the local SMBH mass function, and the local $M_\mathrm{BH}$--$M_\mathrm{bulge}$ relation.
Further validation includes  examination of the Eddington ratio distribution \citep{shirakata_edd_dis}, which showed consistent alignment with observational data. Moreover, \citet{shirakata_soltan} confirmed the model's agreement with the {So\l tan's} argument. Complementary studies by \citet{oogi2016} revealed qualitative concordance with observational results in quasar clustering, and subsequent work by \citet{oogi2023} using more comprehensive $N$-body simulations further explored AGN luminosity functions and cosmic variance.

The merging histories of DM halos are extracted from the micro-Uchuu simulation \citep{uchuu21}, which uses $640^3$ particles, a box length of $100 \, h^{-1} \, \mathrm{Mpc}$, a softening length of $4.27 \, h^{-1} \, \mathrm{kpc}$, and a particle mass resolution of $3.27 \times 10^8 \, h^{-1} \, \mathrm{M}_\odot$. These simulations provide precise information about the structure and evolution of DM halos, which are critical for modeling galaxy formation.
The cosmological parameters used in the simulation are consistent with the Planck 2015 results \citep{planck15}:
\[
\Omega_0 = 0.31, \quad \lambda_0 = 0.69, \quad \Omega_b = 0.048, \quad \sigma_8 = 0.83, \quad n_s = 0.96,
\]
and the Hubble constant,$H_0 = 100\, h \, \text{km} \, \text{s}^{-1} \, \text{Mpc}^{-1}$ where $h = 0.68$.
All results in this study are based on the micro-Uchuu simulation. The model parameters have been recalibrated specifically for the Uchuu merger trees, following the method described in \citet{oogi2023}, ensuring consistency between the model and simulation.

\subsection{Baryonic processes}\label{BaryonicProcesses}
\noindent
Here, we describe the baryonic processes relevant to our study, including gas cooling, star formation, feedback mechanisms, and quenching processes. The full details of the model can be found in \citet{nu2GC} and \citet{nu2GCagn}.

Unless otherwise stated, we adopt the default parameter set of \citet{oogi2023} throughout this paper, in which the model parameters were recalibrated for the Uchuu simulations.

\subsubsection{Gas cooling}\label{GasCooling }
The mass fraction of the baryonic matter in a DM halo is given as $f_\mathrm{b} = \Omega_\mathrm{b} / \Omega_\mathrm{m}$ before the cosmic reionization.
After the reionization ($z < 9$), we reduce the baryon fraction {in small halos} as suggested by \citet{ogt08} (see more details in \citet{nu2GC}).
Initially, all baryonic matter in a halo is diffuse hot gas. To calculate the cold gas mass, we estimate the cooling radius, $r_\mathrm{cool}(t)$, as the radius where the cooling time is equal to the time elapsed since the halo formation. The accretion radius, $r_\mathrm{acc}$, is the minimum of $r_\mathrm{cool}$, the free-fall radius, and the virial radius of the halo. Gas within {$r_\mathrm{acc}(t) < r \le r_\mathrm{acc}(t + \Delta t)$, where $\Delta t$ is the current timestep of the merger tree,} cools and accretes onto the cold gas disk, which increases the cold gas mass, $M_\mathrm{cold}$. The cooling timescale, $t_\mathrm{cool}$, is calculated as:
\begin{equation}
t_{\mathrm{cool}}(r) = \frac{3}{2} \frac{\rho_{\mathrm{hot}}(r)}{\mu m_p } \frac{k_B T_{\mathrm{vir}}}{n_e^2(r) \Lambda(T_{\mathrm{vir}}, Z_{\mathrm{hot}})},
\label{eq:1}
\end{equation}
where $\mu$, $m_p$, $k_B$, and $n_e$ denote the {mean} molecular weight, proton mass, Boltzmann constant, and electron number density, respectively.
For the density of hot gas, $\rho_{\text{hot}}(r)$, we assume the cored isolated isothermal profile \citep{nu2GC}.
The cooling function, $\Lambda(T, Z)$, is calculated using the \citet{sd93} cooling functions assuming the hot gas has the virial temperature of the DM halo. The hot gas metallicity $Z_\mathrm{hot}$ is calculated by solving the chemical evolution as we describe later.

When the mass of the DM halo has doubled since its formation, we update the halo formation epoch and add the gas mass proportional to the increased DM halo mass to the hot halo gas mass, $M_\mathrm{hot}$. We also assume that the reheated gas by supernova (SN) feedback (see Section \ref{StarFormation}) returns to the halo at this epoch.
Using the new virial temperature, new virial radius, and new hot gas mass, we construct a new hot gas profile with $r_\mathrm{cool} = 0$.

\subsubsection{Star formation and supernova feedback}\label{StarFormation}
\noindent
The increase rate in stellar mass, $\dot{M}_{\mathrm{*}}$, is given by the star formation rate, $\Psi(t)$, multiplied by the locked-up mass fraction, $\alpha = 0.52$ \citep{chabrier03}:
\begin{equation}
\dot{M}_{\mathrm{*}} = \alpha \Psi(t),
\label{eq:2}
\end{equation}
where the star formation rate is given by the cold gas mass and the star formation timescale, $\tau_\mathrm{*}$:
\begin{equation}
\Psi(t) = \frac{M_{\mathrm{cold}}}{\tau_{\mathrm{*}}}.
\label{eq:3}
\end{equation}
In our model, the star formation timescale in a disk has dependence on the
disk rotation velocity, $V_{\mathrm{d}}$, as well as the dynamical timescale of the disk, $\tau_{\mathrm{d}}$, to explain the cold gas mass fraction in dwarf galaxies \citep{nu2GC}:
\begin{equation}
\tau_{\mathrm{*}} = \epsilon_{\mathrm{*}}^{-1} \tau_{\mathrm{d}} \left[ 1 + \left( \frac{V_{\mathrm{d}}}{V_{\mathrm{*}}} \right)^{-\alpha_{\mathrm{*}}} \right],
\label{eq:4}
\end{equation}
where $\epsilon_{\mathrm{*}}, V_{\mathrm{*}}, \alpha_{\mathrm{*}}$ are set to 0.46, 197 $\mathrm{km}\,\mathrm{s}^{-1}$, and 2.14, respectively, and $\tau_\mathrm{d}$ is given by the effective radius of the disk, $R_\mathrm{d}$, and $V_\mathrm{d}$ as $\tau_\mathrm{d} \equiv R_\mathrm{d}/V_\mathrm{d}$.
During starburst events initiated by either galaxy mergers or disk instabilities, we employ a star formation timescale much shorter than the merger tree's timestep, ensuring that all available gas is consumed within a single timestep.

When stars form, they release energy and metals into the interstellar medium (ISM) through SN explosions. The energy released by SNe reheats the cold gas,
ejecting a fraction of the cold gas from the galaxy.
The reheated gas mass, $\dot{M}_{\mathrm{reheat}}$, is given by
\begin{equation}
\dot{M}_{\mathrm{reheat}} = \beta \Psi(t),
\label{eq:5}
\end{equation}
where
\begin{equation}
  \beta(V_d) = \left( \frac{V_d}{V_{\mathrm{hot}}} \right)^{-\alpha_{\mathrm{hot}}}.
\label{eq:6}
\end{equation}
In our fiducial model, we employ $\alpha_{\mathrm{hot}} = 3.92$ and $V_{\mathrm{hot}} = 121.64 \, \mathrm{km\,s^{-1}}$.
The reheated gas {will} only be available for cooling when the halo mass doubles.

As a consequence of the star formation and feedback, the cold gas mass evolves as:
\begin{equation}
\dot{M}_{\mathrm{cold}} = -(\alpha + \beta) \Psi(t).
\label{eq:7}
\end{equation}
The metal contents of the cold gas and hot gas change as:
\begin{equation}
\frac{\dm}{\dm t} (M_{\mathrm{cold}} Z_{\mathrm{cold}}) = \left[ p - (\alpha + \beta) Z_\mathrm{cold}  \right] \Psi(t)
\label{eq:8}
\end{equation}
and
\begin{equation}
\frac{\dm}{\dm t} (M_{\mathrm{hot}} Z_{\mathrm{hot}}) = \beta Z_\mathrm{cold} \Psi(t),
\label{eq:9}
\end{equation}
where $p = 1.68 \, Z_\odot$ is the chemical yield.
In the case of a starburst, $\alpha + \beta$ in Eqs.~(\ref{eq:7}) and (\ref{eq:8}) should be replaced by $\alpha + \beta + f_\mathrm{BH}$, where $f_\mathrm{BH} \Psi(t)$ represents the accretion rate onto the central SMBH, and the parameter $f_\mathrm{BH}$ is fixed at 0.02 to match observations \citep{nu2GCagn}.
We model chemical enrichment from star formation and supernova (SN) explosions based on \citet{1992A&A...264..105M}. Specifically, we adopt the instantaneous recycling approximation (IRA) for Type~II SNe, assuming that metals are immediately returned to and uniformly mixed within both the cold and hot gas phases upon star formation (see Eqs.~(\ref{eq:8}) and (\ref{eq:9})). This approximation effectively shortens the metal enrichment timescale and may lead to a slight overestimation of metallicities. We also neglect contributions from Type~Ia SNe.
\subsection{Starbursts triggered by galaxy mergers and disk instabilities}

A starburst can quench star formation by consuming all the cold gas in a galaxy. In our model, starbursts occur via two primary channels: galaxy mergers and disk instabilities \citep{nu2GCagn}. We briefly describe our modeling below.

\subsubsection{Galaxy mergers}

Galaxies can merge through two mechanisms in our model: dynamical friction (central–satellite mergers) and random collisions (satellite–satellite mergers). The timescales of these processes are given in \citet{nu2GC}.

When two galaxies merge, the more massive one is considered the primary, and the less massive one is the secondary. Stars and cold gas from the secondary galaxy are added to the bulge of the primary galaxy. Additionally, the primary's bulge acquires stars and cold gas from its own disk. The fractions of stars and cold gas transferred from the disk to the bulge depend on the merger mass ratio, cold gas fraction, and bulge-to-total mass ratio (see \cite{nu2GCagn} for details). When the merger mass ratio exceeds $f_{\rm major} = 0.89$, the disk of the primary galaxy is completely destroyed, and all the cold gas is consumed by a starburst. In this case, the merger remnant is a pure bulge galaxy without cold gas. If there is no subsequent supply of cold gas after the major merger, the galaxy becomes passive.

\subsubsection{Disk instabilities}

Another starburst-triggering mechanism is disk instability. When a disk becomes gravitationally unstable, fractions of stars and cold gas in the disk are added to the bulge, and the cold gas supplied to the bulge is consumed by a starburst. This occurs when the following condition is satisfied:
\begin{equation}
\frac{V_{\rm max}}{\sqrt{G M_{\rm disk}/r_{\rm ds}}} < \epsilon_{\rm DI,crit},
\end{equation}
where $V_\mathrm{max}$ is the maximum rotational velocity of the dark matter halo hosting the galaxy, and $r_\mathrm{ds}$ is the disk scale length, defined as $r_\mathrm{ds} = (1/\sqrt{2}) \langle \lambda_\mathrm{H} \rangle R_\mathrm{init}$. Here, $R_\mathrm{init}$ is the initial radius of the hot gas sphere, and $\langle \lambda_\mathrm{H} \rangle$ is the mean value of the dimensionless spin parameter of the halo. We adopt $\langle \lambda_\mathrm{H} \rangle = 0.042$ \citep{bett07}. We use $\epsilon_\mathrm{DI,crit} = 0.75$ to reproduce the observed star formation rate density of the universe \citep{nu2GCagn}.

Although the fractions of stars and cold gas transferred to the bulge during disk instability depend on factors such as the bulge-to-disk mass ratio and gas fraction, the amount of cold gas available for a starburst is very small in our model \citep{shimizu24}. Therefore, disk instabilities are expected to have little impact on quenching star formation.

\subsection{Quenching mechanisms}\label{Quenching mechanism}
\noindent
Our model explicitly incorporates two quenching processes. The first is strangulation. When two or more halos merge, the central galaxy of the most massive halo becomes the central galaxy of the newly formed common halo, while the others become satellite galaxies. Any gas that has not yet cooled is assumed to be shock-heated to the virial temperature of the common halo. This hot gas is assumed to cool only onto the central galaxy, so star formation in satellite galaxies continues only until their existing cold gas reservoirs are depleted.

Although this assumption is commonly used in SA models \citep{kwg93, cole94}, several recent models adopt a gradual removal of the hot gas reservoir in satellite galaxies \citep{fon08, Henriques2015, gp18, Cora2018, lagos2024}, motivated by results from hydrodynamic simulations \citep{Kawata08, McCarthy2008}. We will revisit the impact of this more realistic treatment of hot gas stripping on the metallicity differences between passive and star-forming galaxies in Section~\ref{sec:gradual}.

The second quenching process is radio-mode AGN feedback \citep{bower06, croton06}. Gas cooling in a halo is suppressed when the cooling time at the cooling radius exceeds the dynamical time and the central supermassive black hole (SMBH) is sufficiently massive \citep{nu2GCagn}. Since this mechanism primarily operates in massive halos, where the cooling time is longer than the dynamical time, it is not a dominant factor in this study, as we do not focus on galaxies with stellar masses exceeding $M_* > 10^{11} \, \mathrm{M}_\odot$.

It is worth noting that we do not explicitly model the so-called quasar-mode AGN feedback. While this omission is common in SA models that include AGN feedback \citep{bower06, croton06}, some models explicitly incorporate quasar-mode AGN feedback \citep{Somerville2008, Menci2023}. This mode may quench star formation at different stellar mass scales than radio-mode feedback. We will discuss its potential effects on metallicity differences in Section~\ref{sec:qso}.

\section{Results}

\noindent
Throughout this paper, we classify model galaxies as passive if their specific star formation rates fall below $10^{11} \, \mathrm{yr}^{-1}$ following \citet{okamoto17}. We have verified that these galaxies consistently populate the red sequence across all our models. All remaining galaxies are classified as star-forming.

Although luminosity-weighted metallicity is more suitable for comparison with observations, we adopt the simpler mass-weighted metallicity to better isolate the mechanisms driving metallicity enhancement in passive galaxies. We have confirmed that using $r$-band luminosity-weighted metallicity does not alter our conclusions, though it is systematically higher than mass-weighted metallicity in star-forming galaxies.

\begin{figure}[h!]
  {\centering
    \includegraphics[width=8.5cm]{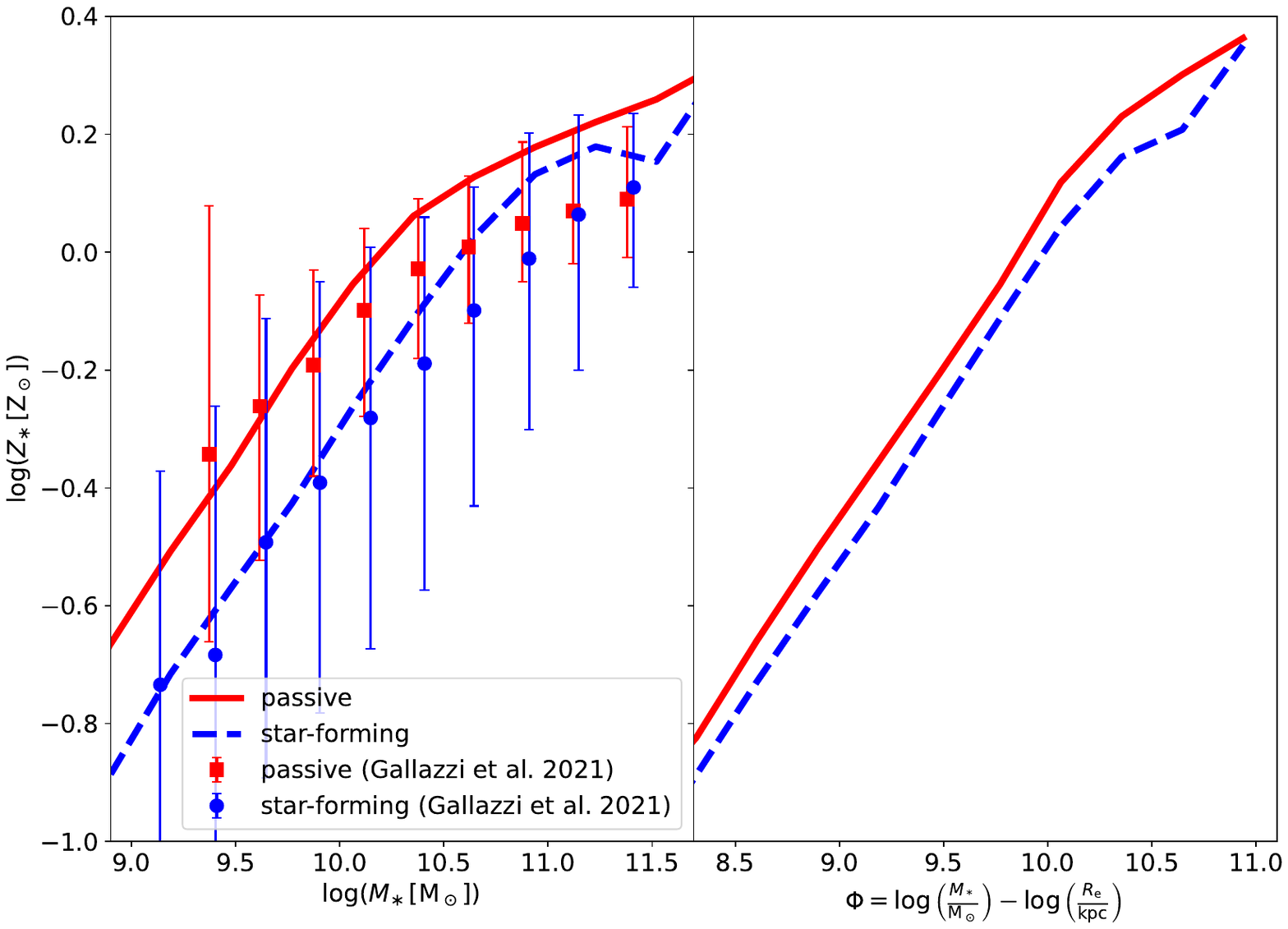}
  }
  \caption{{\it Left}: Stellar MZRs for passive and star-forming galaxies in the fiducial model. The red solid and blue dashed lines represent the medians of the model predictions for passive and star-forming galaxies, respectively. The red filled circle and blue filled squares represent the MZRs of the SDSS galaxies \citep{gallazzi21} compiled by \citet{Fontanot2020}. for quiescent and star-forming galaxies, respectively.  {\it Right}: The same as the left panel, but the metallicities are plotted against the potential, $\Phi = \log(M_*/R_\mathrm{e})$.
  }
  \label{fig:fig1}
\end{figure}
In the left panel of Fig.~\ref{fig:fig1}, we present the stellar MZRs for passive and star-forming galaxies in the fiducial model.
We find that passive galaxies exhibit significantly higher stellar metallicities than star-forming galaxies, in agreement with observations. 
{
For the massive galaxies, the metallicity gap becomes much smaller than for the less massive galaxies with $M_* \lesssim 10^{11} \, \mathrm{M}_\odot$. 
This result is also qualitatively consistent with observations. 
Given that more massive galaxies have shorter star formation timescales (Eq.~(\ref{eq:4})), this trend supports the idea that strangulation is the primary driver of the metallicity difference.
}
The right panel of Fig.~\ref{fig:fig1} shows stellar metallicities as a function of gravitational potential, defined as $\Phi = \log(M_*/\mathrm{M}_\odot) - \log({R_\mathrm{e}}/\mathrm{kpc})$. The effective radius, $R_\mathrm{e}$, is calculated assuming a de Vaucouleurs profile for the bulge and an exponential profile for the disk.
We find that the stellar metallicities of passive galaxies are only slightly higher than those of star-forming galaxies at fixed $\Phi$, consistent with the results of \citet{vaughan22}.

To investigate the origin of the metallicity difference between the two galaxy populations, we examine how the star formation timescale, $\tau_*$, influences their metallicities. Since strangulation timescales are governed by $\tau_*$, we vary the star formation efficiency, $\epsilon_*$, which controls $\tau_*$ as defined in Eq.~(\ref{eq:4}).
We present results for two models: a ``high SFE'' model with $\epsilon_* = 1.0$ and a ``low SFE'' model with $\epsilon_* = 0.1$.

\begin{figure}[h!]
  {\centering
   \includegraphics[width=8.5cm]{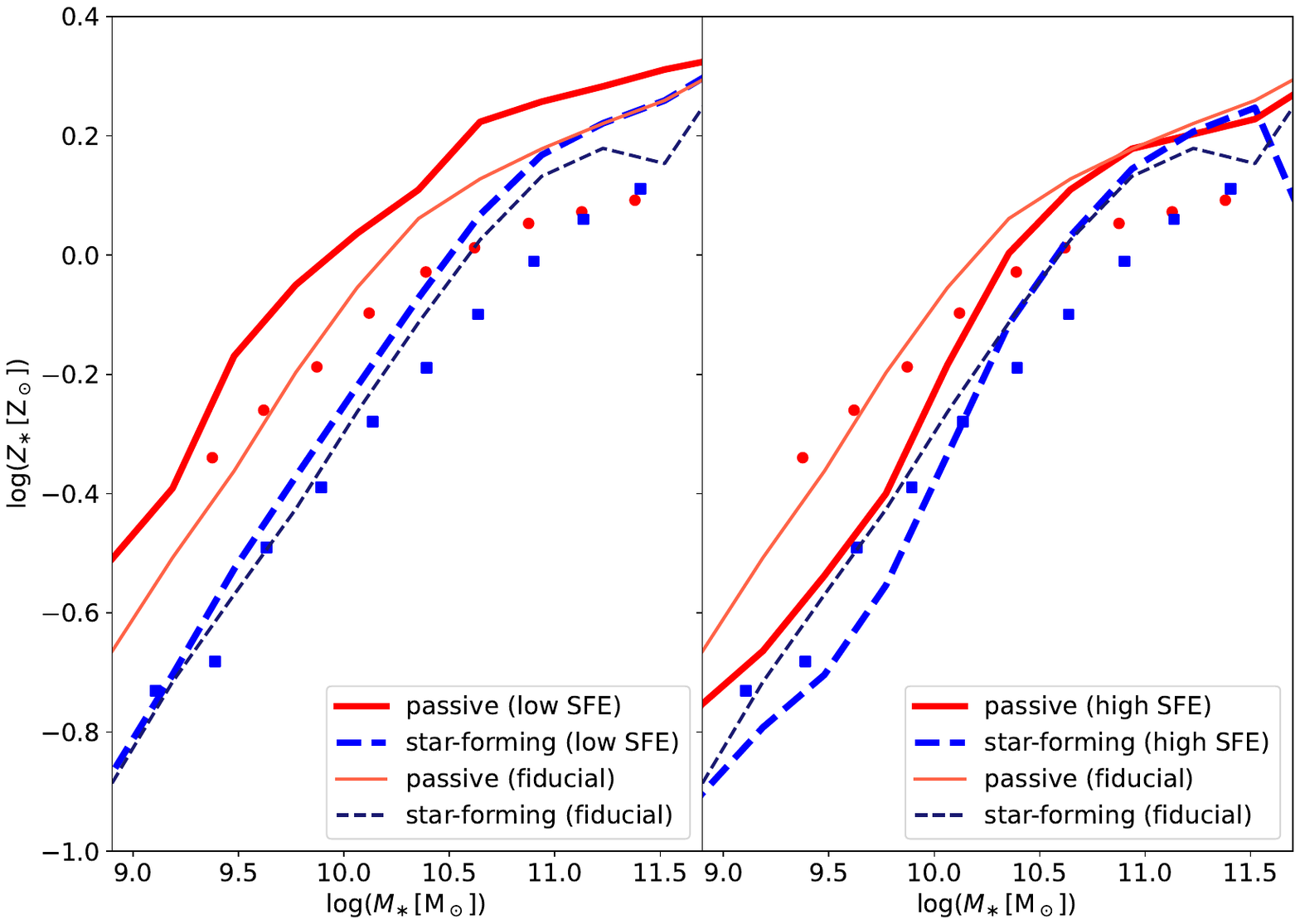}
  }
  \caption{The left and right panels show the stellar MZRs for passive and star-forming galaxies in the low SFE and high SFE models, respectively.
  The predictions by the fiducial model is also shown by the thin lines.
 
  }
  \label{fig:fig2}

\end{figure}
In the left and right panels of Fig.~\ref{fig:fig2}, we present the stellar MZRs for passive and star-forming galaxies in the low SFE and high SFE models, respectively.
We find that the metallicity difference between passive and star-forming galaxies increases with {lower} star formation efficiency (longer $\tau_*$) and decreases with {higher star formation efficiency}.
This finding supports the strangulation hypothesis, since a longer star formation timescale leads to more dilution of the cold gas metallicity for star-forming galaxies.

\begin{figure}[h!]
  {\centering
  \includegraphics[width=8.5cm]{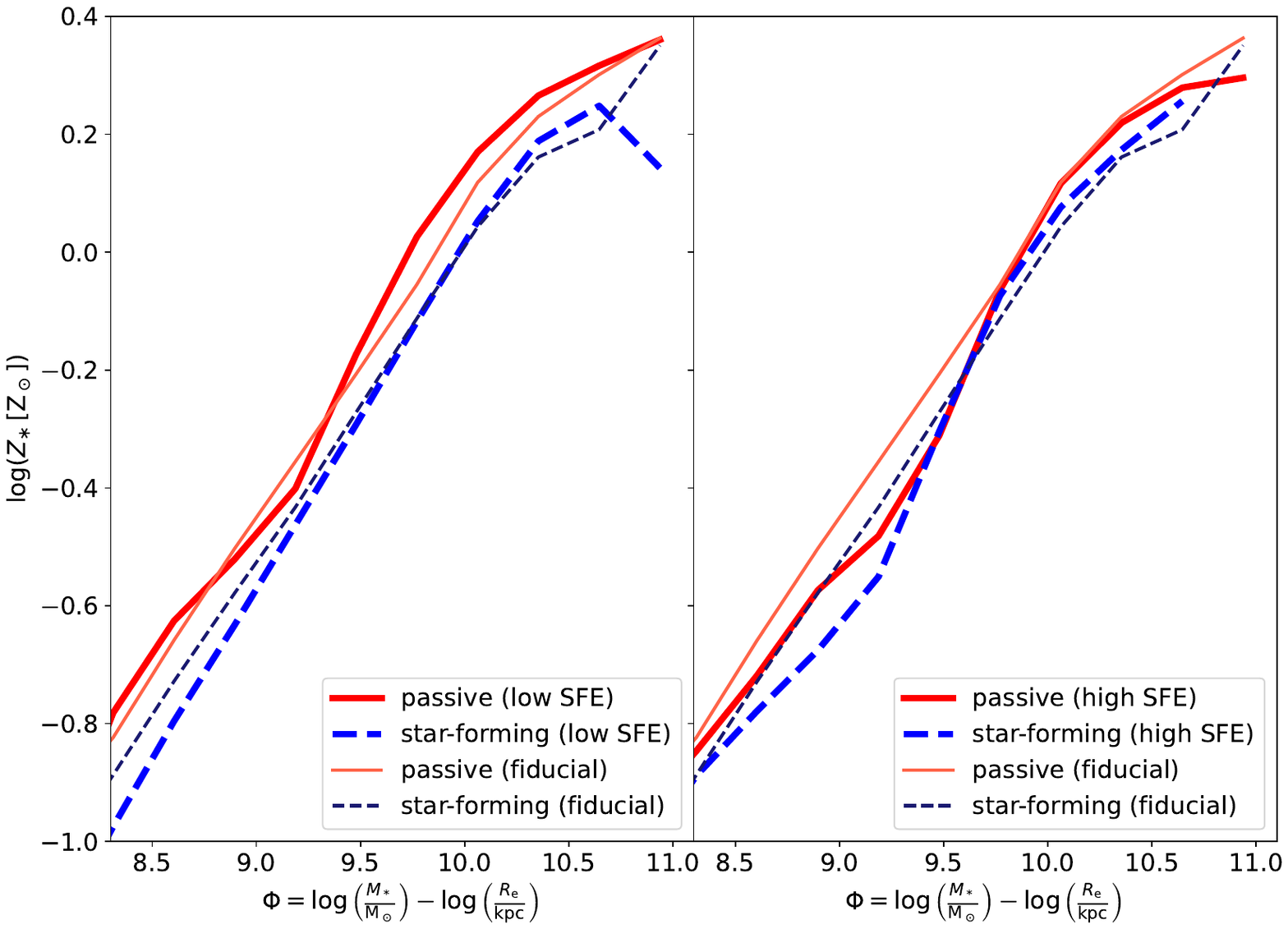}
  }
  \caption{The left and right panels show the stellar metallicities for passive and star-forming galaxies against $\Phi$ in the low SFE and high SFE models, respectively.
  The predictions by the fiducial model are also shown by the thin lines.
 
  }
  \label{fig:fig3}
\end{figure}
In Fig.~\ref{fig:fig3}, we present the stellar metallicities of passive and star-forming galaxies as a function of $\Phi$.
Stellar metallicity is less sensitive to the adopted star formation timescale when plotted against $\Phi$ than as a function of stellar mass.
This suggests that stellar metallicity is primarily regulated by stellar feedback.
However, metallicity differences between star-forming and passive galaxies persist even at fixed $\Phi$ and vary with star formation timescale, indicating that dilution-free self-enrichment during strangulation also plays a role.

%
\begin{figure}[h!]
  {\centering
    \includegraphics[width=8.5cm]{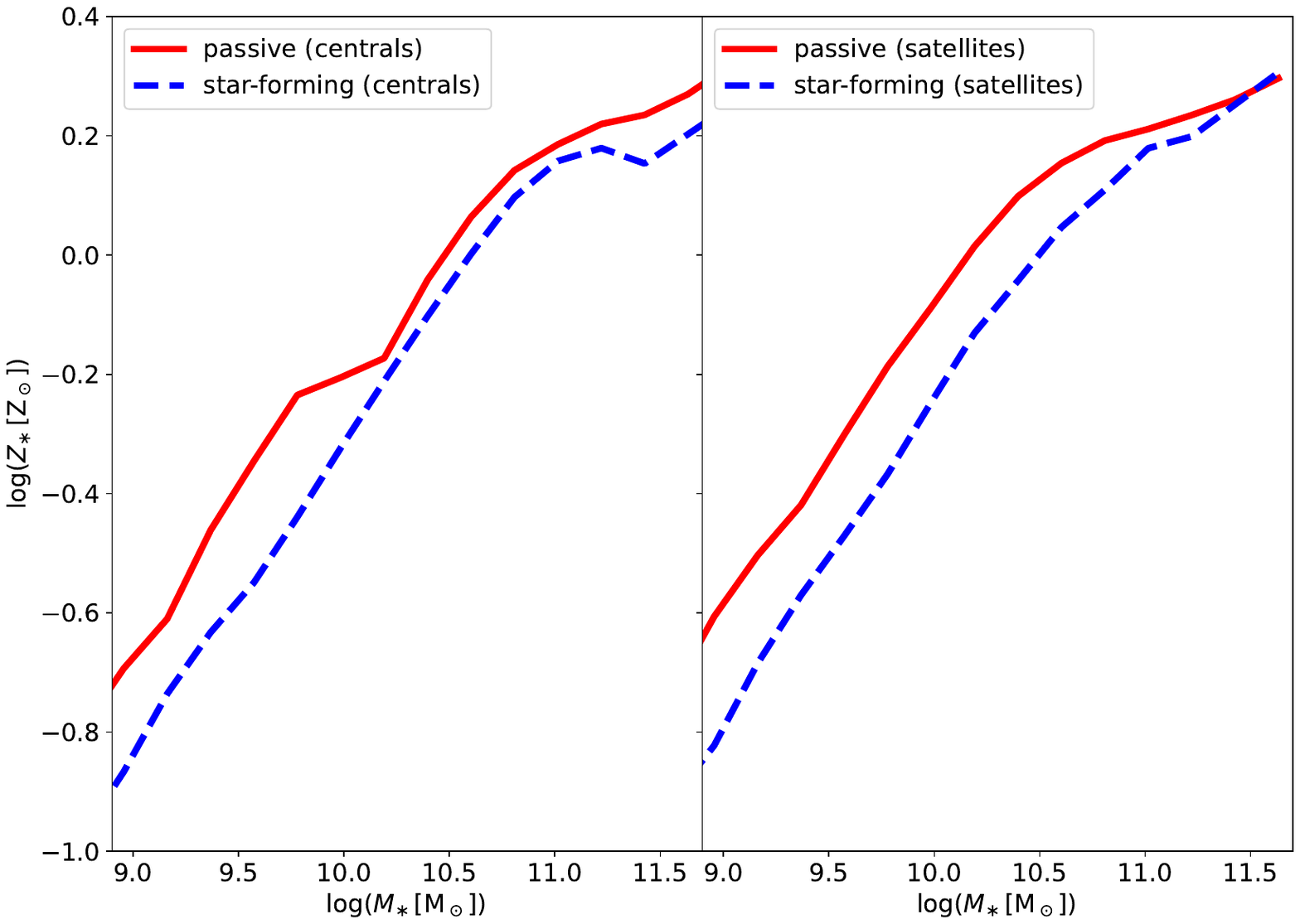}
  }
  \caption{The left and right panels show the stellar MZRs for passive and star-forming galaxies for the central and satellite galaxies in the fiducial model, respectively.
 }
  \label{fig:fig4}

\end{figure}
To investigate whether strangulation explains the metallicity difference more directly, we analyze the stellar MZRs separately for central and satellite galaxies, since strangulation only affects satellite galaxies.
Fig.~\ref{fig:fig4} shows the stellar MZRs for central and satellite galaxies in the fiducial model.
We find that for galaxies with $M_* \gtrsim 10^{10} \, \mathrm{M}_\odot$, the metallicity difference between passive and star-forming galaxies is much smaller for central galaxies compared with the total population, while it remains significant for satellite galaxies. This finding supports the strangulation hypothesis. 
We note that the star-forming satellite galaxies are in the process of being quenched by strangulation; they have not yet fully consumed their cold gas.

Central passive galaxies exhibit a distinct metallicity trend depending on stellar mass. For massive systems ($M_* \gtrsim 10^{10} \, \mathrm{M}_\odot$), their metallicities are only slightly higher than those of star-forming centrals of comparable mass. In contrast, low-mass passive centrals ($M_* \lesssim 10^{10} \, \mathrm{M}_\odot$) show significantly enhanced metallicities.

In our model, central galaxies become passive under specific conditions: either when their host halos grow too slowly, leading to the depletion or ejection of available gas through star formation and stellar feedback before the halos can double in mass, or when radio-mode AGN feedback is triggered. Although radio-mode AGN feedback can produce effects similar to strangulation by shutting down gas cooling, we focus here on the scenario in which central galaxies are quenched in slowly growing halos, since AGN feedback is not a dominant mechanism in the mass range of interest ($M_* \lesssim 10^{11} \, \msun$).

In halos hosting relatively massive galaxies ($10^{10} \lesssim M_*/\msun \lesssim 10^{11}$), cooling times are longer than or comparable to the dynamical time, but radio-mode AGN feedback remains ineffective\footnote{The effect of radio-mode feedback is clearly seen for the most massive central galaxies ($M_* \gtrsim 10^{11}\,\msun$).}. Star formation, therefore, continues using cold gas, whose metallicity remains diluted due to the ongoing accretion of low-metallicity gas from the hot halo. In this mass range, most passive central galaxies are formed via starbursts that consume the remaining cold gas. As a result, their stellar metallicities are only slightly higher than those of their star-forming counterparts.

In contrast, halos hosting galaxies with $M_* \lesssim 10^{10}~\msun$ have very short cooling times, and the hot halo gas cools rapidly and efficiently \citep{rees_ostriker1977, correa18}. In such cases, chemical evolution proceeds without significant metal dilution of the cold gas, closely resembling the strangulation scenario.

\begin{figure}[h!]
  {\centering
    \includegraphics[width=8.5cm]{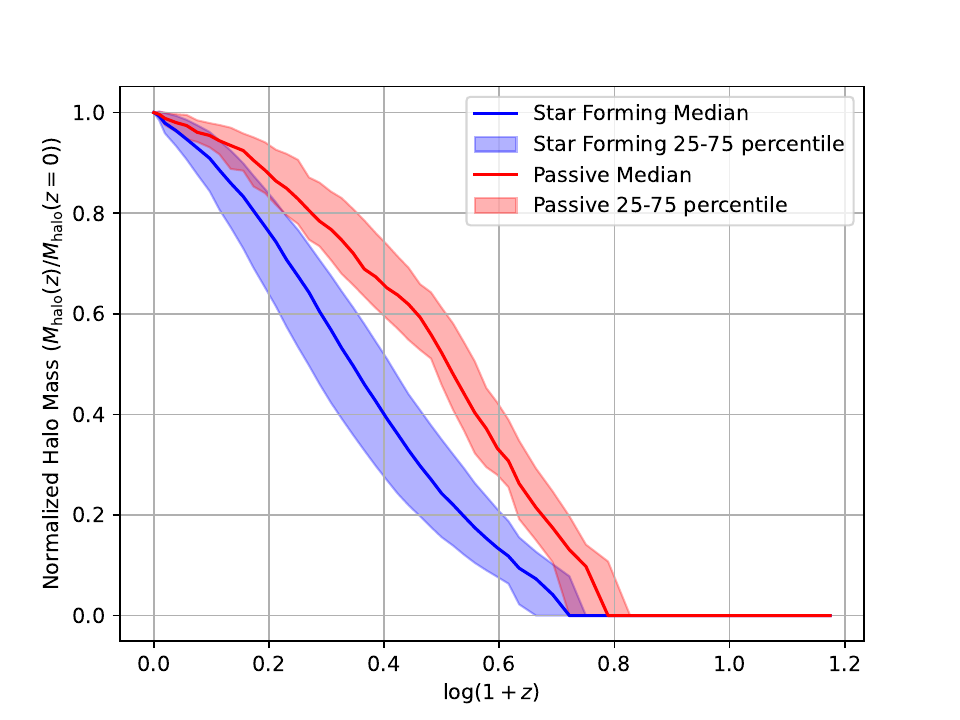}
  }
  \caption{
  Normalized halo mass evolution as a function of redshift for both star-forming and passive galaxies whose stellar mass at $z = 0$ is $\sim 10^{9.7} \mathrm{M}_\odot$. The vertical axis represents the normalized halo mass $M_{\mathrm{halo}}(z) / M_{\mathrm{halo}}(z=0)$, and the horizontal axis denotes redshift. The blue line and shaded region represent the median and 25\%-75\% percentile range for star-forming galaxies, while the red line and shaded region correspond to passive galaxies. 
  }
  \label{fig:fig5}
\end{figure}
If our hypothesis is correct, the host halos of the central passive dwarf galaxies should exhibit significantly higher formation redshifts compared to those of the central star-forming dwarf galaxies. To test this, we examine the formation histories of the host halos of central dwarf galaxies in Fig.~\ref{fig:fig5}. We randomly select 10 halos for both passive and star-forming galaxies with stellar masses of $\sim 10^{9.7} \, \mathrm{M}_\odot$ at $z = 0$. The results show that the host halos of the central passive dwarf galaxies have higher formation redshifts and do not exhibit significant evolution at low redshifts, supporting our hypothesis.

The quenching induced by slow halo growth is likely an artificial effect resulting from the model of gas accretion onto halos. In reality, there is no physical justification for restricting gas accretion to instances where the halo mass doubles. However, this result reinforces the conclusion that passive galaxies exhibit higher stellar metallicities than star-forming galaxies of the same stellar mass when quenching occurs through a process analogous to strangulation.
Nonetheless, as shown in Fig.~\ref{fig:fig4}, this artificial effect is not the primary driver of metallicity enhancement in passive galaxies within our model, at least for $M_* \gtrsim 10^{10} \, \msun$.

\section{Discussion}\label{Discussion}

\noindent
We have demonstrated that the observed metallicity difference between passive and star-forming galaxies can be explained if most passive galaxies are quenched through strangulation. 
This raises the question of why previous studies using cosmological simulations and SA models have failed to reproduce this metallicity difference, despite including strangulation as a quenching mechanism. Here, we discuss a potential explanation and present models that fail to produce the observed metallicity difference.

We also examine the impact of physical processes not included in our current model, such as gradual hot gas stripping and quasar-mode AGN feedback, on the metallicity gap.

\subsection{Schmidt law as a star formation model}
\noindent
Star formation is often modeled using the Schmidt law, which posits that the star formation timescale is proportional to the dynamical time. \citet{Fontanot2020} adopted this star formation model, as described in {\citet{delucia04}}. Compared with our star formation model defined by Eq.~(\ref{eq:4}), this approach could assume significantly shorter star formation timescales for dwarf galaxies {with $V_\mathrm{d} < V_*$}.
To explore how the choice of star formation models influences the metallicities of passive and star-forming galaxies, we present results obtained with a model where the star formation timescale is given by
{
\begin{equation}
\tau_* = \epsilon_*^{-1} \tau_{\mathrm{d}},
\label{eq:10}
\end{equation}
instead of Eq.~(\ref{eq:4}). We set $\epsilon_* = 0.23$ 
}
to ensure that a galaxy with $V_\mathrm{d} = V_* = 197 \, \mathrm{km \, s}^{-1}$ has the same star formation timescale as in the fiducial model.

\begin{figure}[h!]
  {\centering
    \includegraphics[width=8.5cm]{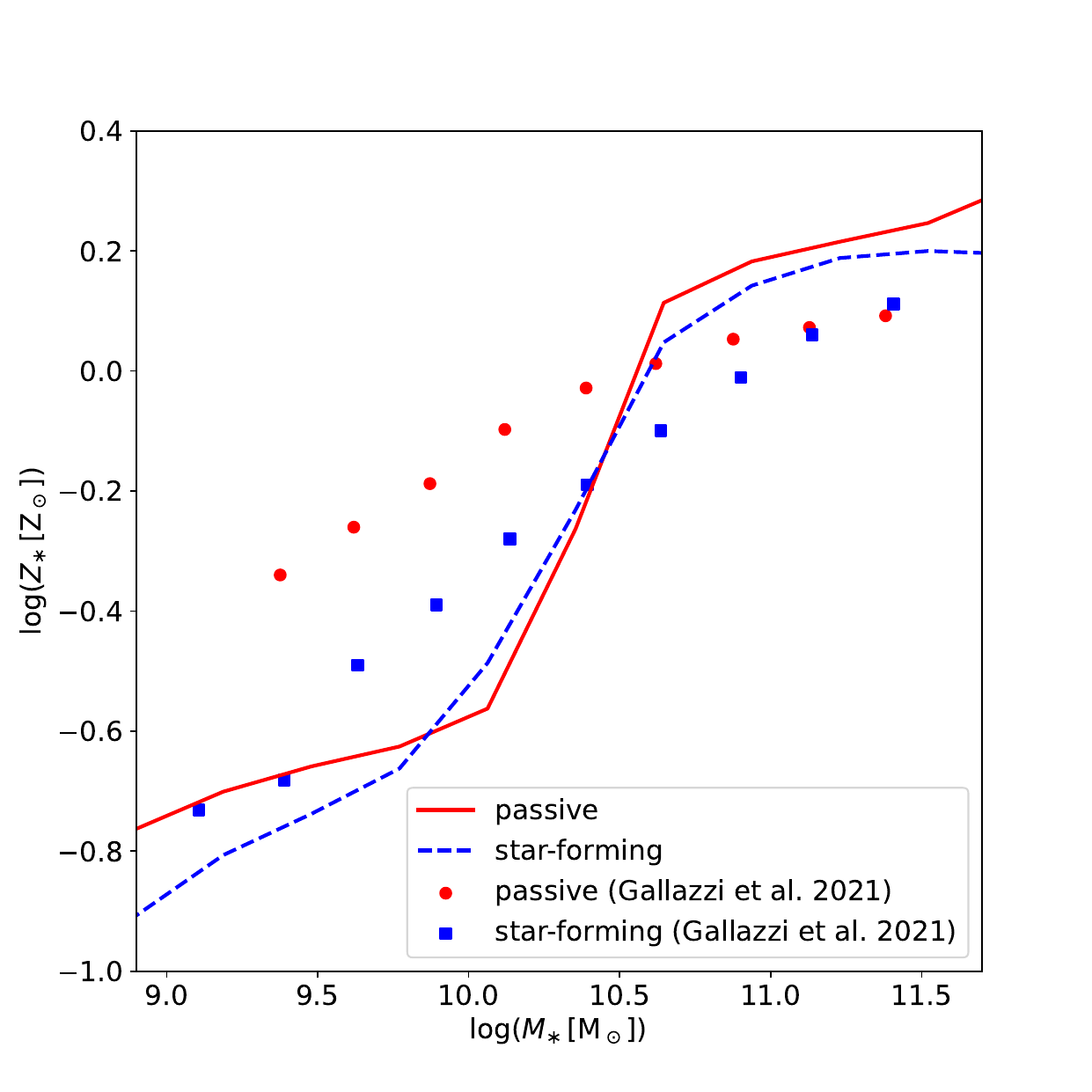}
  }
  \caption{
    The same as the left panel of Fig.~\ref{fig:fig1} but for the model in which the star formation timescale is proportional to the dynamical time. 
  }
  \label{fig:fig6}
\end{figure}
Fig.~\ref{fig:fig6} shows the MZRs for both quiescent and star-forming galaxies in this model. The observed trend of enhanced metallicities in quiescent galaxies at fixed stellar mass is absent in this result. 
This suggests that adopting longer star formation timescales in low-mass galaxies ($M_* \lesssim 10^{11} \, \mathrm{M}_\odot$) is crucial to reproducing the observed metallicity difference. 
We note that this model is not expected to match the observations used to constrain the fiducial model parameters \citep{nu2GCagn} since the model for star formation is significantly modified without adjusting other physical processes. 

The inability of the cosmological simulations analyzed in \citet{okamoto17} to reproduce the observed metallicity difference between star-forming and passive galaxies is probably due to the same issue: the star formation timescales in low-mass galaxies are too short.
As shown by \citet{illustris} and \citet{okamoto14}, the mean stellar ages of their simulated dwarf galaxies are systematically older than observed.
To address this issue in cosmological hydrodynamical simulations, it would be necessary to introduce physical processes that preferentially extend the star formation timescales in low-mass galaxies.

\subsection{Instantaneous removal of cold gas}
\noindent 
We have demonstrated that the observed metallicity difference between star-forming and passive galaxies can be explained if most passive galaxies were quenched through strangulation over sufficiently long timescales. Since the quenching timescale in strangulation directly corresponds to the star formation timescale, it remains unclear whether quenching duration or star formation duration is the more fundamental factor.
To investigate this further, we consider a model with instantaneous quenching, where both cold and hot gas are stripped from a galaxy as soon as it becomes a satellite. All other physical processes remain identical to those in the fiducial model.
We do not attempt to model ram-pressure stripping of cold gas, as galaxies are generally preprocessed by strangulation before experiencing this effect \citep{on04}.
\begin{figure}[h!]
  {\centering
     \includegraphics[width=8.5cm]{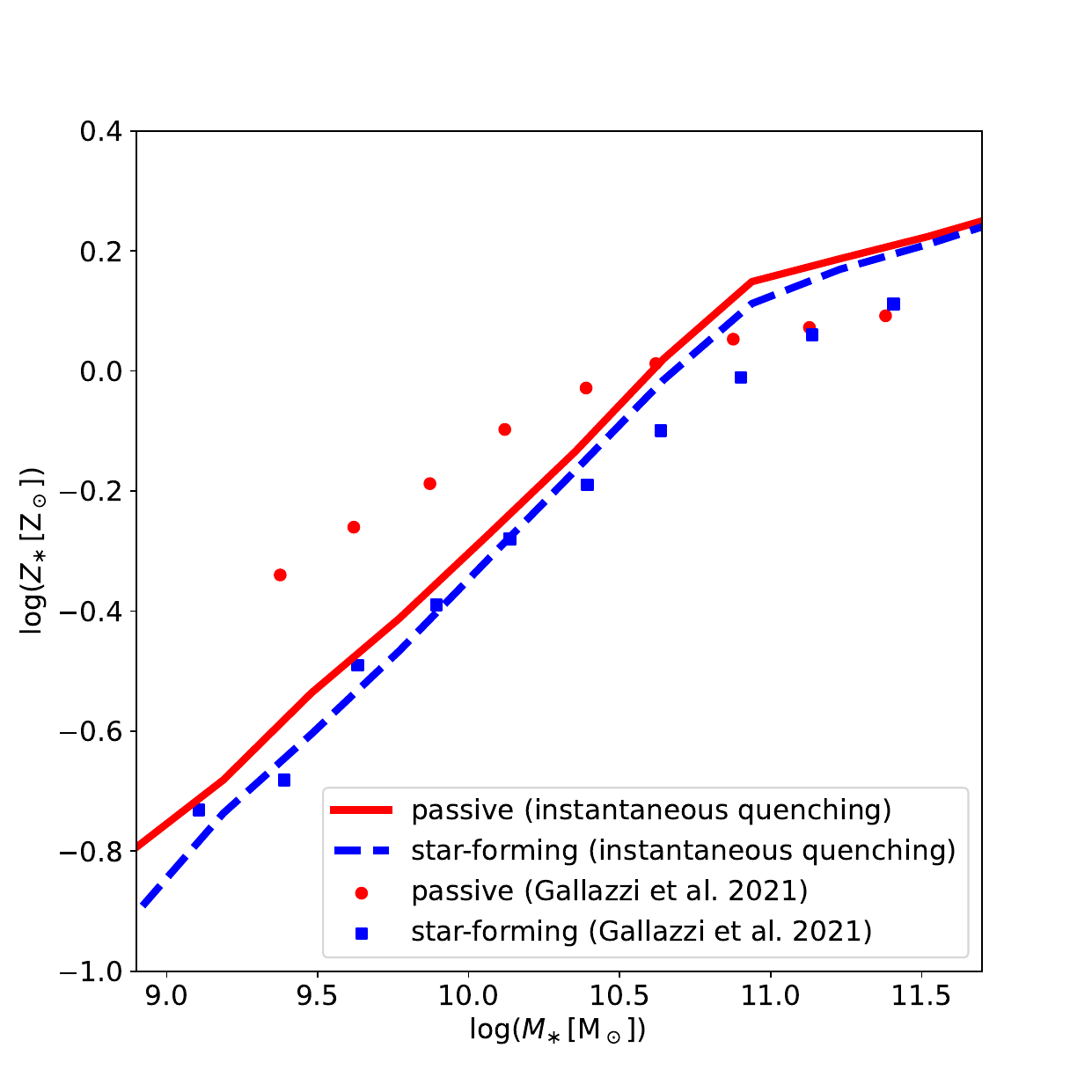}
  }
  \caption{
    The same as Fig.~\ref{fig:fig1} but instantaneous cold gas removal for satellite galaxies is assumed. 
    }
  \label{fig:fig7}
\end{figure}
Fig.~\ref{fig:fig7} illustrates the metallicities of star-forming and passive galaxies under the assumption of instantaneous quenching. In this model, the metallicities of passive galaxies are nearly identical to those of star-forming galaxies, whether compared at the same stellar mass or at the same $\Phi$.

Passive galaxies exhibit only slightly higher metallicities than star-forming galaxies at a given stellar mass. 
Because instantaneous quenching removes the most metal-enriched gas, passive galaxies cannot reach metallicities as high as those observed in the fiducial model.
The similar metallicities at a fixed $\Phi$ suggest that stellar metallicity is primarily governed by feedback processes. However, quenching mechanisms can still influence the metallicities of passive galaxies. Unlike stellar feedback, instantaneous cold gas stripping removes cold gas without further star formation, leading passive galaxies to exhibit slightly lower metallicities than their star-forming counterparts at fixed $\Phi$.

These results highlight the importance of strangulation in producing the significantly higher stellar metallicities observed in passive galaxies  star-forming galaxies of the same stellar mass.

\subsection{Gradual hot gas stripping} \label{sec:gradual}

While strangulation is modeled in our framework as the instantaneous removal of the hot gas reservoir when a galaxy becomes a satellite, numerical simulations suggest that hot gas is removed more gradually \citep{Kawata08, McCarthy2008}. Accordingly, several studies have incorporated gradual stripping processes into their models \citep{fon08, Henriques2015, gp18, Cora2018, lagos2024}. Although the stripping timescale depends on the specific implementation, it is typically on the order of $\sim$~Gyr.

Here, we investigate the impact of gradual hot gas stripping from satellites on the metallicity gap by applying a constant stripping timescale to all satellites, modeled as:
\begin{equation}
M_\mathrm{hot}(t + \Delta t) = M_\mathrm{hot}(t)\exp\left(-\frac{\Delta t}{\tau_\mathrm{strip}}\right),
\end{equation}
where $\Delta t$ is the time interval between two consecutive snapshots, and $\tau_\mathrm{strip}$ is the stripping timescale. We assume that hot gas is stripped from the outer envelope of the satellite’s hot gaseous halo.

\begin{figure}[h!]
  {\centering
    \includegraphics[width=8.5cm]{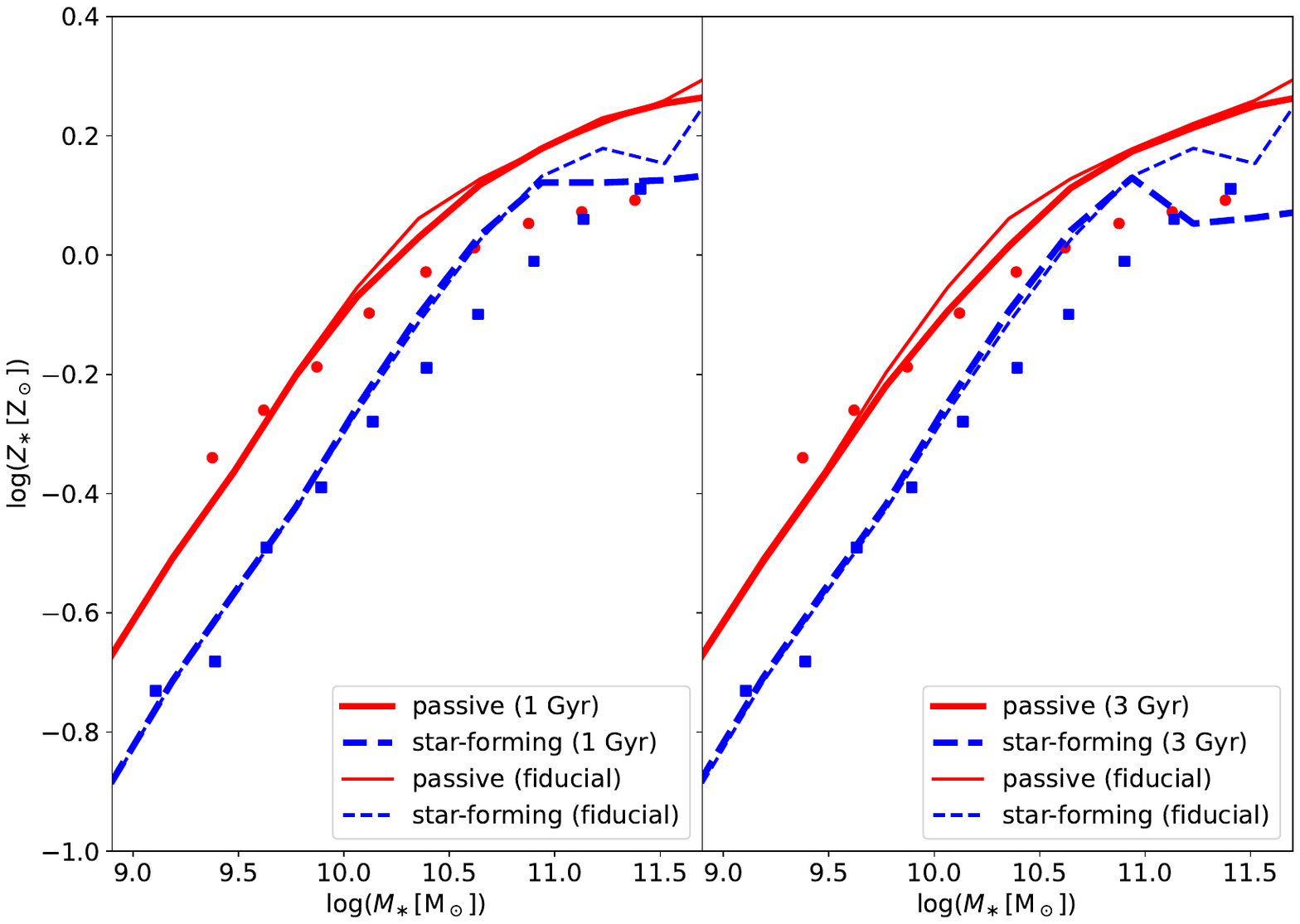}
  }
  \caption{Stellar MZRs for passive and star-forming galaxies in models with gradual hot gas stripping. We assume $\tau_\mathrm{strip} = 1$~Gyr and 3~Gyr in the left and right panels, respectively. Line styles follow the same convention as in previous figures.
  }
  \label{fig:gradual}
\end{figure}
In Fig.~\ref{fig:gradual}, we present the results for $\tau_\mathrm{strip} = 1$~Gyr and 3~Gyr in the left and right panels, respectively. While a longer stripping timescale increases the fraction of star-forming satellites, the metallicity difference between passive and star-forming galaxies remains nearly unchanged. This is because the metallicity enhancement, caused by the absence of dilution, only begins after the hot gas is mostly depleted. Thus, the magnitude of the metallicity gap continues to be primarily governed by the star formation timescale, even when gradual hot gas stripping is introduced.\footnote{Strictly speaking, the gap becomes slightly narrower for longer $\tau_\mathrm{strip}$ due to ongoing metal dilution during the stripping phase. Note also that modifying the hot gas stripping prescription for satellites can indirectly affect central galaxies by altering the total hot gas content in their halos.}

\subsection{Quasar-mode AGN feedback} \label{sec:qso}

While our model includes only radio-mode AGN feedback, some studies incorporate quasar-mode AGN feedback either in addition to or instead of radio-mode feedback \citep{Somerville2008, Menci2023}.  
Quasar-mode feedback suppresses star formation primarily through galaxy-scale winds driven by luminous, rapid black hole accretion, which is typically associated with starbursts. 
As such, it generally plays a significant role during gas-rich major mergers.

\citet{vaughan22} showed that the observed metallicities of passive and star-forming galaxies, as functions of both stellar mass and potential, can be reproduced if more compact galaxies at fixed stellar mass have a higher probability of being quenched. 
If gas-rich mergers tend to form compact bulges, the metallicity gap could potentially be explained when quasar-mode feedback is responsible for quenching relatively low-mass galaxies with $M_* \lesssim 10^{11}~\msun$. 
However, exploring this possibility is beyond the scope of the present study.

\section{Conclusion}
\noindent 
Using the $\nu^2$GC semi-analytic model, we have examined the underlying mechanisms responsible for the observed stellar metallicity differences between star-forming and passive galaxies at a fixed stellar mass. Our fiducial model succeeded in qualitatively reproducing the metallicity difference between these two populations as a function of stellar mass and gravitational potential.
Our analysis reveals that strangulation---a quenching mechanism that halts cold gas supply while allowing existing gas to form stars---plays a crucial role in generating this metallicity difference. When star formation occurs over sufficiently extended timescales, the termination of cold gas inflow prevents dilution of the metal-enriched gas reservoir. This process enables passive galaxies to achieve higher stellar metallicities compared to star-forming galaxies of equivalent stellar mass.
We have established a direct relationship between the star formation timescale and the magnitude of the metallicity difference: longer timescales produce more pronounced metallicity gaps between passive and star-forming populations. Importantly, our results demonstrate that instantaneous cold gas stripping, unlike strangulation, fails to generate the observed metallicity differential, highlighting the time-dependent nature of chemical enrichment in galactic evolution.

We also examined why previous studies using SA models and cosmological simulations have failed to reproduce the observed metallicity difference. We tested a model in which the star formation timescale is proportional to the dynamical time---a common approach in SA models \citep{delucia04} and naturally expected in hydrodynamic simulations. This formulation leads to star formation timescales that are too short for low-mass galaxies, resulting in insufficient metallicity growth in passive galaxies during strangulation. To address this issue, hydrodynamic simulations would need to include additional physical processes that preferentially extend the star formation timescales in low-mass galaxies, aligning their predictions with observations.

While our fiducial model successfully reproduces the observed metallicity difference between star-forming and passive galaxies, it predicts significant metallicity enhancement in central dwarf passive galaxies, which should not experience strangulation. Our analysis reveals that these galaxies reside in DM halos with short cooling times and high-formation redshifts. Under our assumption that gas cannot accrete onto a halo until it doubles its mass since its last formation epoch, galaxies in such rapidly cooling halos experience conditions analogous to strangulation until the halos double their masses. We speculate that this discontinuous treatment of gas accretion contributes to the overproduction of red dwarf galaxies in our model \citep{nu2GC}. To address this, a more realistic treatment of continuous gas accretion alongside the mass growth of host DM halos would be essential.

Our study confirms that the observed metallicity difference supports strangulation as the primary quenching mechanism. Since the timescale of strangulation is determined by the star formation timescale, this metallicity difference serves as a valuable constraint for star formation models in both simulations and SA models.

\begin{ack}
This study is supported by JSPS/MEXT KAKENHI Grants (21H04496, 20H05861, 19H01931, 20H01950, 18H05437, 23K03460, 21H05449, and 20K22360) and JST SPRING, Grant Number JPMJSP2119. 
We thank Instituto de Astrofisica de Andalucia (IAA-CSIC), Centro de Supercomputacion de Galicia (CESGA) and the Spanish academic and research network (RedIRIS) in Spain for hosting Uchuu DR1 and DR2 in the Skies \& Universes site for cosmological simulations. The Uchuu simulations were carried out on Aterui II supercomputer at Center for Computational Astrophysics, CfCA, of National Astronomical Observatory of Japan, and the K computer at the RIKEN Advanced Institute for Computational Science. The Uchuu DR1 and DR2 effort has made use of the skun@IAA\_RedIRIS and skun6@IAA computer facilities managed by the IAA-CSIC in Spain (MICINN EU-Feder grant EQC2018-004366-P).
\end{ack}

\bibliography{okamoto_ref}
\end{document}